\documentclass[aip,chaos,reprint,groupedaddress]{revtex4-1}

\usepackage{natbib,dcolumn,graphicx}
\usepackage{lineno,todonotes,amssymb}

\begin{document}
\title{Complex network based techniques to identify extreme events and (sudden) transitions in spatio-temporal systems}

\author{Norbert Marwan}
\email{marwan@pik-potsdam.de}
\author{J\"urgen Kurths}
\altaffiliation[Also at ]{Humboldt Universit\"at zu Berlin, Institut f\"ur Physik}
\affiliation{Potsdam Institute for Climate Impact Research, 14412 Potsdam, Germany}

\date{\today}

\begin{abstract}
We present here two promising techniques for the application of the complex network approach to 
continuous spatio-temporal systems that have been developed in the last decade and show large potential
for future application and development of complex systems analysis. First, we discuss the transforming of
a time series from such systems to a complex network. The natural approach is to calculate the recurrence matrix
and interpret such as the adjacency matrix of an associated complex network, called recurrence network. Using 
complex network measures, such as transitivity coefficient, we demonstrate that this approach is very efficient for 
identifying qualitative transitions in observational data, e.g., when analyzing paleoclimate regime transitions. 
Second, we demonstrate the use of directed spatial networks
constructed from spatio-temporal measurements of such systems that can be derived from the synchronized-in-time 
occurrence of extreme events in different spatial regions. Although there are many possibilities to investigate such
spatial networks, we present here the new measure of network divergence and how it can be used to develop a prediction
scheme of extreme rainfall events.
\end{abstract}


\maketitle 

\begin{quotation}
In the last decades various powerful techniques have been developed in nonlinear dynamics for the study of continuous spatio-temporal
dynamic systems. Typically they are based on different discretization methods in space and time. Here, we discuss an unconventional 
approach based on complex networks for the investigation of such systems. In contrast to well-known examples of complex networks, 
such as social ensembles, neural networks, or power grids, where the nodes are clearly defined by humans, neurons or power 
generating stations, here the first step of the complex network approach can be interpreted as a very flexible way to discretize a 
continuous system, or to identify a backbone underlying the continuous system. This enables us to use in the next steps the rich 
variety of methods from complex network theory even for the analysis of continuous systems. Based on this approach we treat 
two basic problems in high-dimensional nonlinear dynamics: (i) uncovering regime shifts and (ii) prediction of extreme events. 
We propose appropriate techniques for both by combining recurrence with networks resp.~synchronization with extreme events. 
The potential of this approach is demonstrated here for the Earth system. In a first example we show that important main regime shifts 
of the East Asian Monsoon during the last 3 Million years can be identified from paleoclimate proxy records. In a second 
example we analyze recent satellite data from the tropical rainfall measurement mission (TRMM) and use the network
divergence for developing an efficient prediction scheme for extreme precipitation events in the eastern Central Andes.
\end{quotation}

\section{Introduction}
Climate is as the brain a highly complex and high-dimensional system; both systems have a lot of 
joint properties, but there are also important distinctions. Understanding the mechanisms of
climatic processes on all temporal and spatial scales is very difficult and even impossible in near future, but 
crucial for weather forecasts or assessment of long-term climate changes.
A data-based investigation of the climate system is related with several challenges, in particular
non-stationarity (e.g., abrupt vs.~slow changes), high-dimensionality, 
non-Gaussian distributed data (e.g., extreme events), natural vs.~anthropogenic influences, etc.

A basic first step in data-based studies of such a complex high-dimensional 
system is reducing the dimensionality.
The most widely used method for this is a decomposition into a very finite number of 
Empirical Orthogonal Functions (EOFs).
This approach also allows to identify main spatial patterns, such as large circulation patterns or 
teleconnections. However, 
the basis of the EOF approach is the covariance matrix, thus, only capturing the first two statistical moments 
and demanding for certain strict properties of the data, e.g., Gaussian distribution and stationarity \cite{monahan2009}. 
Even more obvious are constraints due to event-like data, 
as typical for rainfall and extreme events, or limitations by nonlinear interrelations. 
Moreover, the found EOFs do not undoubtedly coincide with
typical climate phenomena \cite{dommenget2002}.

Modern measurement techniques has allowed in the last decades to extend our knowledge 
into the past, leading to paleoclimatology. However, these data generate further challenges: 
Dating uncertainties and
irregularly sampled time series are problems that limit the direct application of standard methods.

An alternative and novel approach for the study of different aspects of the climate system is related to the
progress in complex networks science in the past quarter century. At a first glance it might appear surprising that the complex network
approach can be used to analyse a continuous system as the climate and in particular to identify
spatio-temporal patterns in climate fields or regime shifts in the
paleoclimate. However, the application of complex networks for climate analysis has become a 
lively and quickly progressive field in the last years. Although, this new approach is still in its infancy, 
first results are very promising and have already shown its impressive potential.

In the following we will present two techniques based on complex networks, recurrence networks
(Sec.~\ref{rec_netw}) and event synchronization (Sec.~\ref{clim_netw}), and will 
show how they can be used to uncover 
regime transitions in the paleoclimate by analyzing proxy records (Sec.~\ref{sudden_trans})
and to analyze spatiotemporal 
patterns of extreme rainfall leading to new prediction schemes (Sec.~\ref{clim_netw}).
Finally we summarize the potentials of this non-traditional approach, but discuss also open problems.

\section{Complex networks}

We give here only a few basics on complex networks which will be used later 
(see, e.g., \cite{strogatz2001,newman2003,boccaletti2006} for more detailed reviews on complex network analysis).
A network is a set of nodes and links. We define a network as complex when its topology is highly irregular.
A network can be defined by the adjacency matrix $\mathbf{A}$. For undirected and unweighted 
networks, $\mathbf{A}$ is a binary matrix, just indicating the existence of links between two nodes.
In weighted networks, a link has a weight, i.e., $\mathbf{A}$ consists of real numbers; $\mathbf{A}$
is symmetric for undirected and asymmetric for directed networks.
A network or its components (links, nodes) can be characterized by several measures. Here we
mention only some selected measures. 

The {\it node degree} in unweighted networks is simply the total number of links a node $i$ has and is given by the column sum of
the adjacency matrix $\mathbf{A}$: 
\begin{equation}\label{eq_degree}
k_i = \sum_j^N A_{ji}.
\end{equation}
The distribution of this measure can be used to investigate, e.g., whether a network is scale-free.
On directed networks, we can distinguish between the column-wise and raw-wise sums in Eq.~(\ref{eq_degree})
that give us the in- and out-degree, respectively.
For weighted networks the sum Eq.~(\ref{eq_degree}) becomes the so-called {\it node strength}, and for 
directed and weighted networks, we can consider the {\it in-} and {\it out-strength} 
\begin{equation}
\mathcal S_{i}^{\text{in}} = \sum_{j = 1}^{N} A_{ij}\quad \text{and}\quad \mathcal S_{i}^{\text{out}} = \sum_{j = 1}^{N} A_{ji}.
\label{eq_in_outdegree}
\end{equation}

Another important measure is the  {\it transitivity coefficient}
\begin{equation}\label{eq_trans}
\mathcal{T} =  \frac{\sum_{i,j,k=1}^N A_{j,k}A_{i,j}A_{i,k}}{\sum_{i,j,k=1}^N A_{i,j}A_{i,k} }.
\end{equation}
It measures the probability that the neighbors of a node are connected themselves.

\section{Recurrence Networks -- a time series analysis approach by means of complex networks}
\label{rec_netw}

Analyzing time series by complex networks is a quite new idea that came up in the last decade. 
The generation of a complex network representation of a time series can be done using different approaches,
e.g., by visibility graphs \cite{lacasa2008} or temporal succession of local rank orders \cite{small2013}.
A quite natural approach is to use the {\it recurrence matrix} \cite{marwan2007} of a dynamical system
\begin{equation}\label{eq_rp}
R_{i,j} = \Theta(\varepsilon - ||\vec x(i) - \vec x(j)||), 
\end{equation}
as the adjacency matrix of a complex network \cite{donner2011}:
\begin{equation}\label{eq_rn}
\mathbf{A} = \mathbf{R} - \mathbb{I}
\end{equation}
(with $\mathbb{I}$ the identity matrix, $\Theta$ the Heaviside function,
$\vec{x}(i)$ a state at time $i = 1,\ldots,N$, and $N$ the number of state vectors). 
The recurrence matrix itself has become a basic tool of nonlinear time series analysis and was first
introduced by Eckmann et al.~as recurrence plots that ``are rather easily obtained aids 
for the diagnosis of dynamical systems'' \cite{eckmann87}. Later
this idea was extended by several quantification approaches \cite{zbilut92,marwan2002herz}
leading to ``an active field, with many ramifications we [Eckmann et al. -- ] had not
anticipated'' \cite{marwan2008epjsteditorial}. Fundamental works on such methodological 
developments have been published also in Chaos, e.g., on embedding issues and dynamical invariants
\cite{iwanski98, thiel2004a}, time-delay systems and non-chaotic strange attractors \cite{senthilkumar2008a,ngamga2008a},
heterogenous recurrence analysis \cite{yang2014}, or twin surrogates \cite{romano2009}.

Because of the striking similarity
between the recurrence matrix and the adjacency matrix (i.e., a binary and square matrix), the idea to
identify the recurrence matrix with the adjacency matrix was so obvious that it came 
up almost at the same time (around 2008) within different research groups 
\cite{xu2008,marwan2009b,strozzi2009}. Its main advantage is that 
the resulting recurrence network can be analyzed by the known network measures, 
i.e., further diagnostic tools become available for time series analysis.
In particular the transitivity coefficient $\mathcal{T}$ is appropriate because it 
quantifies the geometry of the phase space trajectory and 
can be used to differentiate between different dynamics (e.g., regular and irregular) \cite{marwan2009b,zou2010}.
It also
allows to define a dimensionality measure \cite{donner2011epjb}, the {\it transitivity dimension}
 \begin{equation}\label{eq_transdim}
D_{\mathcal{T}} = \frac{\log(\mathcal{T})}{log(3/4)}.
\end{equation}

The quantification of the recurrence matrices can also be performed by the recurrence quantification analysis 
(RQA) \cite{marwan2007,webber2015}. In contrast to the network measures which describe the
geometrical properties, the RQA measures characterize dynamical properties of the phase space
trajectory. Therefore, the recurrence network based measures provide complementary information
to the RQA and can, under certain circumstances, give more insights into the system's behavior. 

\begin{figure*}[htbp]
\begin{center}
\includegraphics[width=.7\textwidth]{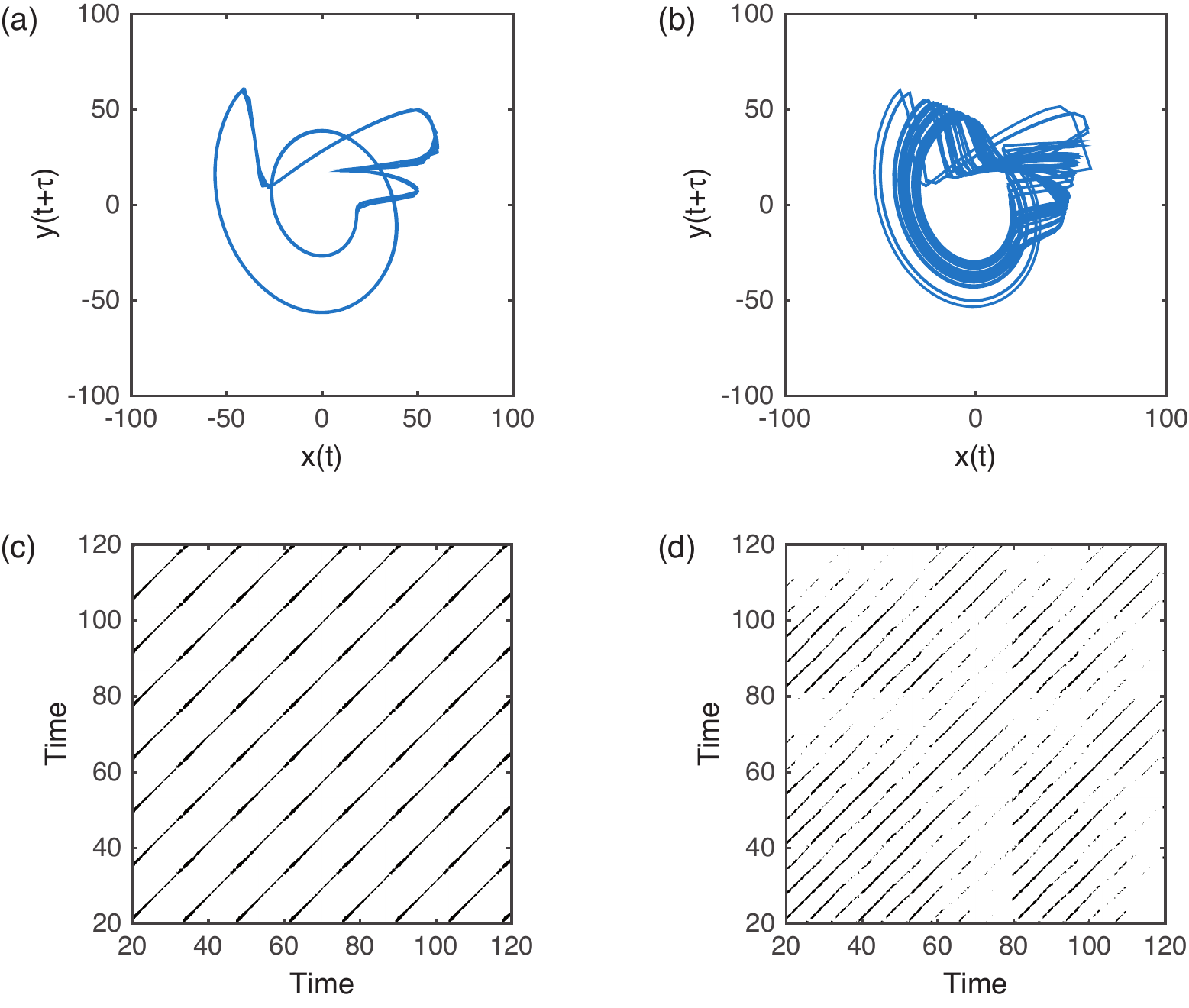}
\caption{(a) Phase space reconstruction of the $x$-component of the R\"ossler system for
$a = 0.245$ and (b) for $a=0.29$. (c) Corresponding recurrence plot for $a = 0.245$, showing periodic 
structures, and (d) for $a=0.29$, showing interrupted diagonal lines. The recurrence plots
are calculated from the $x$-component  using time-delay embedding with $m=4$ and $\tau=17$ (sampling time $\Delta t = 0.1$).}
\label{roessler_RPs}
\end{center}
\end{figure*}

\begin{figure*}[htbp]
\begin{center}
\includegraphics[width=.7\textwidth]{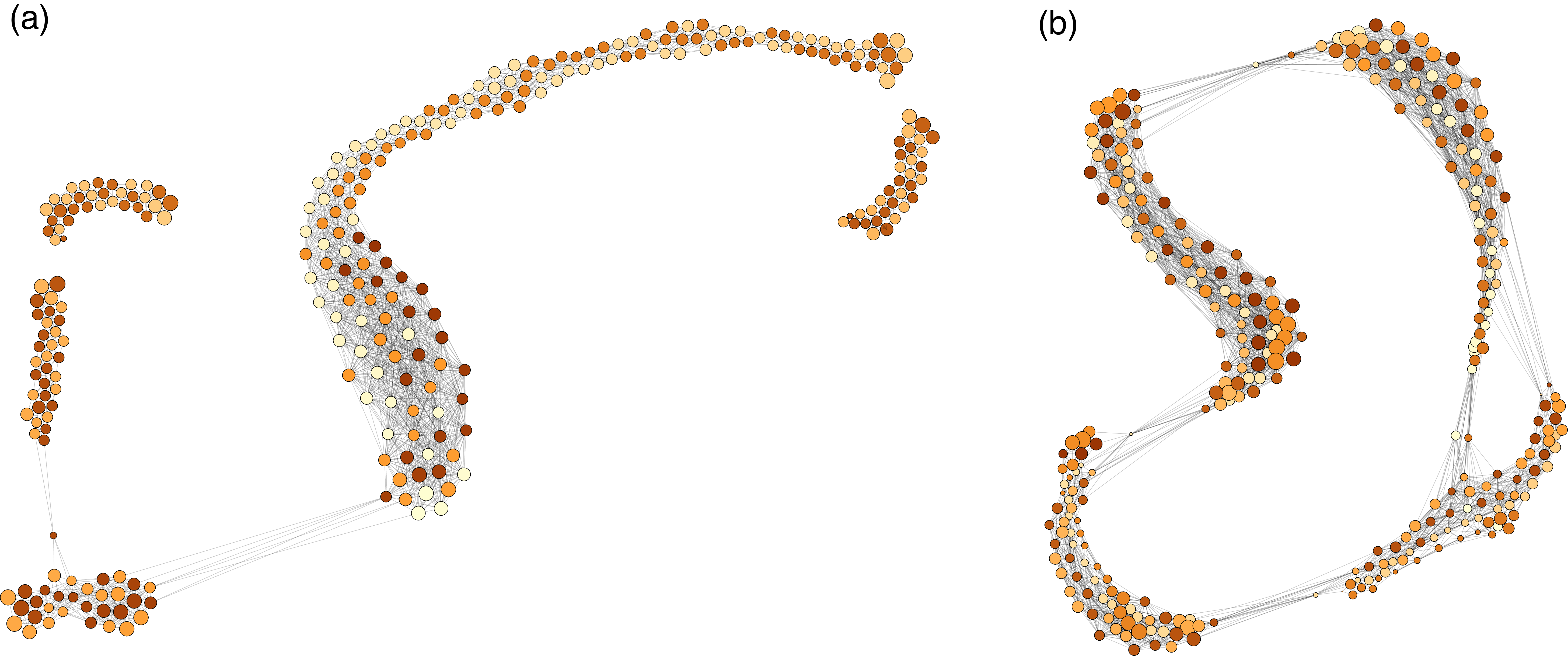}
\caption{Network representation of the phase space reconstruction of the $x$-component of the R\"ossler system for
(a) $a = 0.245$ and (b) $a=0.29$. The network is constructed from the first 300 nodes of the
recurrence matrix shown in Fig.~\ref{roessler_RPs} by a linear repulsion model. Node size and color can be used
to represent selected node properties, here time (node color, the darker the larger $t$) and clustering coefficient
(node size).}
\label{roessler_net}
\end{center}
\end{figure*}

\begin{figure}[htbp]
\begin{center}
\includegraphics[width=1\columnwidth]{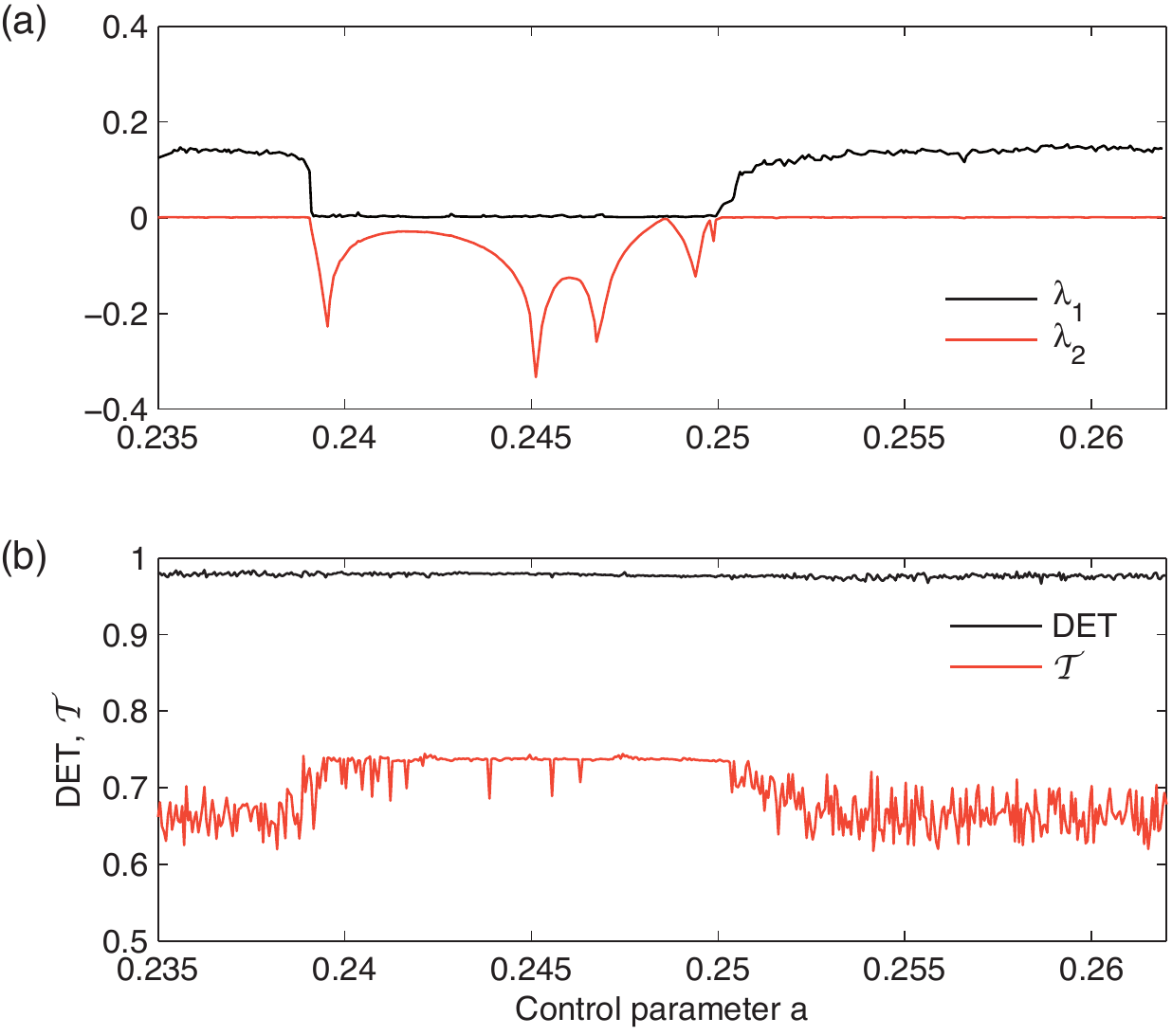}
\caption{(a) The two largest Lyapunov exponents and (b) the recurrence measures DET and $\mathcal{T}$
for the $x$-component of the R\"ossler system for varying control parameter $a$. Within the interval
$a = [0.24 \ 0.25]$, the dynamics is periodic. Whereas $\mathcal{T}$ indicates the periodic behavior
by increased values, it is difficult to detect the periodic window with DET \cite{marwan2011}.}
\label{roessler_RQA}
\end{center}
\end{figure}

As a paradigmatic example, let us consider the R\"ossler system \cite{roessler1976}
\begin{equation} \label{ros_eqs}
\left(\frac{dx}{dt},\,\frac{dy}{dt},\,\frac{dz}{dt}\right) =
\bigl(-y-z,\ x+ay,\ b+z(x - 35)\bigr),
\end{equation}
where we change the parameter $a=b$ in a range where the system shows chaotic and periodic dynamics:
$a = 0.235, \ldots, 0.262$ (Figs.~\ref{roessler_RPs} and \ref{roessler_net}). Between $a=0.24$ and 0.25, the system does not have a
positive Lyapunov exponent and generates periodic behavior (Fig.~\ref{roessler_RQA}(a)).

A frequently used RQA measure for differentiating periodic and chaotic dynamics is 
the ratio of recurrence points that form diagonal lines in the recurrence plot,
called {\it determinism} \cite{trulla96,marwan2007}:
\begin{equation}
DET = \frac{\sum_{l \ge l_{\min}} l P(l)}{\sum_{i,j} R_{i,j}},
\end{equation} 
with $P(l)$ the histogram of line lengths in the recurrence plot.
The idea of this measure is that the length of a diagonal line in the recurrence plot corresponds to the time
the system evolves very similar as during another time. Such repeated similar state evolution 
that is also related to predictability is typical for deterministic systems. In contrast, systems with independent
subsequent values, like white noise, have mostly single points in the recurrence plot. DET is sensitive
to transitions between chaotic and periodic dynamics in maps \cite{trulla96}, but for continuous
systems, such as our R\"ossler example, this measure fails for this task \cite{zou2010,marwan2011}.
For the entire range of the considered $a$ values, it has very high values, close to one (Fig.~\ref{roessler_RQA}(b), Tab.~\ref{tab_results_roessler}).
However, $\mathcal{T}$ shows increased values within the periodic window, close to a value
of $3/4$. 
In general, the network based measures, such as $\mathcal{T}$, 
can add further important aspects in recurrence analysis, in particular for the 
uncovering of sudden changes of the dynamics.

\begin{table}[htbp]
\caption{Typical values of DET and $\mathcal{T}$ for different dynamical
regimes in the R\"ossler system.}\label{tab_results_roessler}
\begin{tabular}{rrrl}
$a$	&DET	&$\mathcal{T}$		&dynamics\\
\hline
0.235	&0.98	&0.66	&chaotic\\
0.245	&0.98	&0.74	&periodic\\
0.260	&0.97	&0.63	&chaotic\\
\hline
\end{tabular}
\end{table}


\section{Identification of sudden transitions in paleoclimate}
\label{sudden_trans}

The recurrence network approach has great potential in different applications in many disciplines.
Using as a classifier, it can help, e.g., to detect serious diseases, such as preeclampsia
\cite{ramirez2013}, to detect epileptic states \cite{lang2013}, or to study multiphase fluid flows \cite{gao2013b}. 
Another important application is to detect critical
transitions in the dynamics \cite{marwan2009b,donges2011d,eroglu2014b}. Such transition detection
is of crucial interest in studying variations of the past climate in order to better understand the climate system in general.

In the following we discuss a typical example from paleoclimate research.
The investigation of relationships between sea surface temperature (SST) and specific
climate responses, like the Asian monsoon system or the thermohaline circulation
in the Atlantic, as well as their regime changes, represents an important scientific challenge for
understanding the global climate system, its mechanisms, and its related variability. 
Its better understanding is of crucial importance as non-linear feedback mechanisms
and tipping points cause high uncertainty and an unpredictable future for humankind\cite{lenton2008,rockstrom2009}.

\begin{figure}[htbp]
\begin{center}
\includegraphics[width=.7\columnwidth]{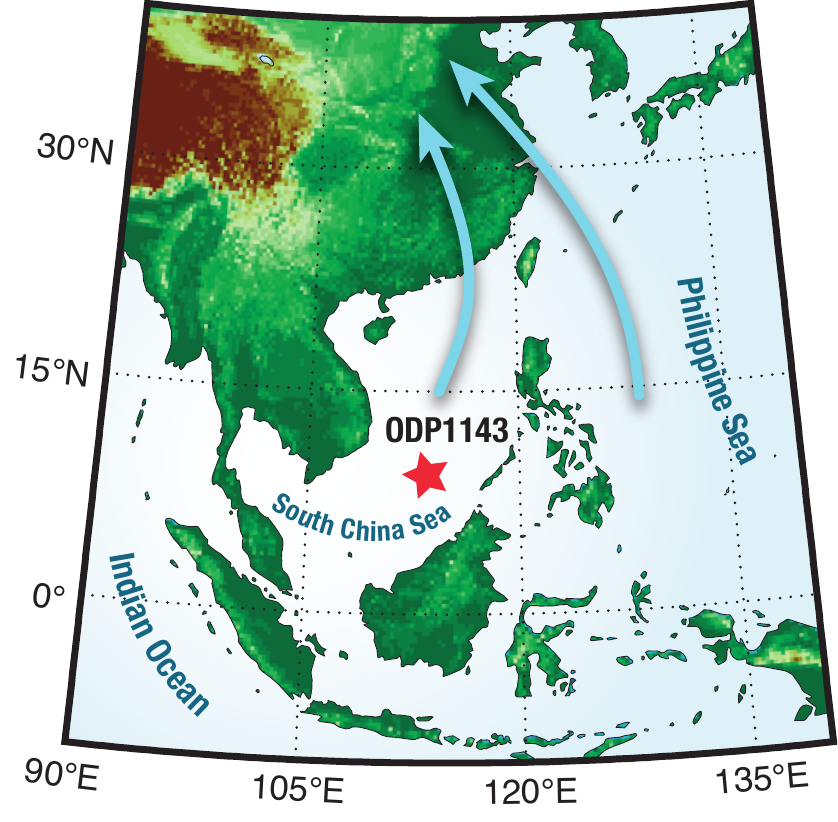}
\caption{Location of the ODP 1143 site in the South China Sea (red star) and
main directions of the East Asian Summer Monsoon (blue arrows).}
\label{map_odp1143}
\end{center}
\end{figure}

\begin{figure}[htbp]
\begin{center}
\includegraphics[width=\columnwidth]{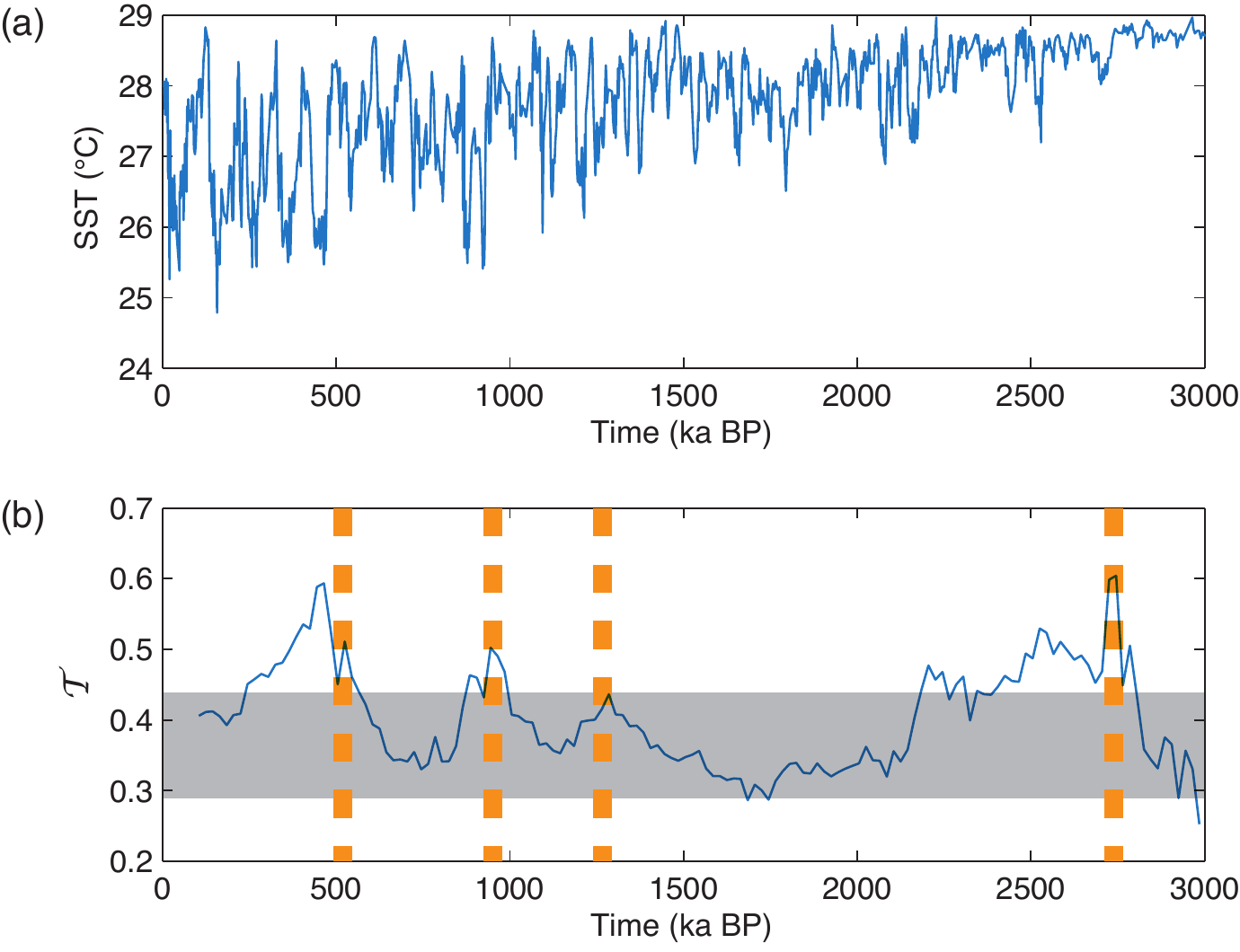}
\caption{(a) Alkenone paleothermometry based SST estimates for the South China Sea and
(b) corresponding transitivity coefficient $\mathcal{T}$. Dashed lines mark major climate shifts discussed in the
text; the gray shading marks the confidence interval of 90\%.}
\label{odp1143_transit}
\end{center}
\end{figure}

In paleoclimatology, different archives are used to reconstruct and study climate
conditions of the past, as lake \cite{marwan2003climdyn} and marine sediments \cite{herbert2010} or
speleothemes \cite{kennett2012}. Alkenone remnants in the organic fraction of
marine sediments, produced
by phytoplankton, can be used to reconstruct SST of the past (alkenone paleothermometry), allowing to study 
the temperature variability of the oceans \cite{herbert2001,li2011epsl}.
Here we will use a SST reconstruction for the South China Sea and the past 3~Ma 
derived from alkenone paleothermometry of the Ocean Drilling Programme (ODP) site 1143\cite{li2011epsl}
(Fig.~\ref{map_odp1143} and \ref{odp1143_transit}(a)). The South China Sea is strongly linked to the East Asian Monsoon system (EAM)
that consists of a winter part with strong winds and a precipitation related summer part.
In general, the East Asian Monsoon is of crucial importance for China's socio-economic behavior, e.g., for agriculture or even 
for public health by its impact on prevalence of trace elements \cite{blazina2014}. The understanding of the
mechanism is, therefore, crucial for learning about the past and the future climate and its impacts.

First we are faced with a typical problem in paleoclimatology because the 
original time series of ODP 1143 is not equally sampled. The sampling time
ranges from 0.2 to 28~ka, with a median of 2.1~ka. If applying standard techniques
(linear methods or classical RQA) then we would first need to interpolate the time series to 
an equidistant time axis. However, when using the recurrence network approach, the 
correct timing of the nodes is not so important (and could even be exchanged without
changing the network properties), because it is characterizing the geometrical structure \cite{donges2011c,donges2011d,feldhoff2012}.

We calculate the recurrence networks and the transitivity coefficient $\mathcal{T}$ for sliding windows of length 410~ka (thus,
with varying number of data points within the windows) and a moving step of 20~ka (Fig.~\ref{odp1143_transit}).
For the phase space reconstruction\cite{packard80} we choose an embedding dimension of $m=6$ (as 
suggested by the false nearest neighbors method\cite{kennel92}).
The selection of the time delay is guided by the auto-correlation function and considered to be constant for
all time windows to be approximately 20~ka (based on median sampling time within one time window). 
The threshold is chosen in such a way to preserve a constant recurrence rate of 7.5\% \cite{marwan2007,donges2011d}.

Moreover, we perform a bootstrapping approach using 1,000 resamplings of the windowed time series for preparing an empirical
test distribution for $\mathcal{T}$. In this real world example, we use a confidence level
of 90\%. As we do not know which kinds of dynamical transition are there,
we will consider both the upper and the lower confidence level.

In the last 3~Ma several major and many smaller climate changes have appeared on regional but
also global scale. Dramatic climate shifts are related with the Milankovich cycles \cite{haug1998,medinaelizalde2005,an2014}
and major changes in ocean circulation patterns \cite{karas2009}. 
Due to a transition towards an obliquity-driven climate variability
with a 41~ka period around 3.0~Ma ago, a period of warm climate has end and the northern hemisphere glaciation
started after 2.8--2.7~Ma \cite{haug1998,herbert2010,an2014}. This transition is very well revealed by the 
significant increase of $\mathcal{T}$ between 2.8 and 2.2~Ma. Based on thorough investigations of loess sediments,
it is known that 1.25~Ma ago the intensity of the winter monsoon of the EAM begun to be strongly coupled to 
global ice-volume change \cite{an2014}. During this time, $\mathcal{T}$ increased (although not reaching significance).
This time also marks the beginning of a transition phase towards glacial-interglacial cycles of 100~ka period
(eccentricity dominated period of the Milankovich cycles). This 100~ka period dominance was well established after 0.6~ka
and is clearly visible by the increased $\mathcal{T}$ between 0.6 and 0.2~ka \cite{sun2010}.
From loess sediments it is also known that the summer monsoon has weakened between 2.0 and 1.5~ka and
around 0.7~ka. During these periods, $\mathcal{T}$ shows lower values than during the previously discussed
periods. The variation of $\mathcal{T}$ confirms the
previous findings of a strong link between the EAM and the  Milankovich cycles, in particular of increased 
and reduced regularity in the climate dynamics (as presented by the SST and for the considered time scale) during 
dominant Milankovich cycles and periods of major climate transitions from one to another regime.

Similar conclusions based on $\mathcal{T}$ have been drawn from dust flux records around the African continent \cite{donges2011d}.
There, it was found that enhanced regular climate dynamics coincides well with lake level high stands in East Africa and, hence, 
indicating that climate regime transitions have triggered human evolution.

\section{Complex networks for spatio-temporal analysis of continuous systems}
\label{clim_netw}

Another important problem in complex systems analysis is the investigation of spatio-temporal dynamics.
In the last decade, this field has also benefited from the complex network approach.
In particular, its application on climate data in order to uncover climate mechanisms
or characteristic spatial patterns and long-range interrelations has drawn attention to
complex networks for spatio-temporal analysis of continuous systems. Complex networks
are an alternative to EOFs and can shed light on different and complimentary aspects than EOFs.
Beginning with the study of Tsonis et al.~in 2004 \cite{tsonis2004}, 
the climate network approach has received more and more interest for spatio-temporal
data analysis\cite{ebertuphoff2012,vandermheen2013,stolbova2014}. 
The idea is to reconstruct a complex network from spatially embedded 
time series (in case of climate, e.g., from a surface air temperature field) by measuring the interrelationship $C_{i,j}$
between these time series. The location of the nodes can be arbitrary (e.g., weather stations when using
instrumental data) or grid points (e.g., when using model or reanalysis data). In unweighted networks, links represent
high correlations between the time series belonging to the nodes, simply considered by applying
a threshold $T$ on the interrelation matrix $\mathbf{C}$ (that could be, e.g., Pearson correlation)
\begin{equation}
A_{i,j} = \left\{
\begin{array}{l l}
  C_{i,j} & \quad \text{if}\quad  C_{i,j} > T,\\
  0 & \quad \text{else}.\\
\end{array} \right.
\label{eq_thresholdC}
\end{equation}
Such networks can be undirected or unweighted (as in Eq.~(\ref{eq_thresholdC})), but also
directed or weighted. 

Within the climate context, such network approach has been applied to study, e.g, 
climate communities \cite{tsonis2004,steinhaeuser2010}, the impact of the El Ni\~no/ Southern Oscillation
\cite{yamasaki2008}, major heat transport pathways and spatio-temporal scales \cite{donges2009,ebertuphoff2012}, 
external and internal atmospheric forcing\cite{deza2014},
to create early warning indicators of critical regime shifts \cite{vandermheen2013}, 
or even for model intercomparison \cite{steinhaeuser2013,feldhoff2014}.
When using Pearson correlation for describing the interrelationships $C_{i,j}$ between the nodes, 
then the node degree is obviously related to the first EOF \cite{donges2015}. Other network
measures, such as betweenness centrality, provide further information that cannot be captured by the EOF analysis \cite{donges2009}.

In general, interrelationships between spatially located time series cannot be considered to be only linear. In order
to examine nonlinear interrelations, information based measures (e.g., mutual information) were suggested for network 
reconstruction \cite{donges2009epjst,barreiro2011,runge2012,deza2013,hlinka2013}. In particular when investigating climatological
or meteorological phenomena, we often face event-like data, such as daily (or hourly) rainfall series or 
extreme events time series. For such kind of data, Spearman rank correlation could be used \cite{carpi2012}.
However, an even more powerful approach for such data is the event synchronization 
approach \cite{quiroga2002,malik2010}. 

Event synchronization was developed to investigate the synchronous activity of the neurons in the brain \cite{quiroga2002}.
It simply
counts the number of temporally coinciding events in two event series $x_1$ and $x_2$
by allowing small deviations between the occurrence of the events, i.e., a dynamical delay between them. 
Let $e_1(m)$ and $e_2(n)$
be the time indices when events appear in $x_1$ and $x_2$ and $m,n = 1,\ldots,l$ the number of 
a specific event ($l$ is the total number of events in the event series). 
The waiting time between an event $m$ in $x_1$ and event $n$ in $x_2$ 
is $d_{1,2}(m,n) = e_1(m) - e_2(n)$. If this waiting time $d_{1,2}(m,n)$ is smaller than some dynamical
delay $\tau(m,n)$, the two events $e_1(m)$ and $e_2(n)$ are considered to occur synchronously.
The dynamical delay $\tau(m,n)$ is the half of the minimal waiting time of subsequent events in
both time series around event $e_1(m)$ and $e_2(n)$ and not larger than a given maximal delay $\tau_{\max}$, i.e., 
\begin{equation}
\tau(m,n) = \min\frac{\lbrace d_{11}(m,m-1), d_{11}(m,m+1), d_{22}(n,n-1), d_{22}(n,n+1), 2 \tau_{\max} \rbrace}{2}.
\label{eq_tau}
\end{equation}
As soon as  $\vert d_{1,2}(m,n)\vert \leq \tau(m,n)$ (or $0 < d_{1,2}(m,n) \leq \tau(m,n)$), we count it as undirected 
(or directed) synchronization of events $e_1(m)$ and $e_2(n)$
\begin{equation}
S(m,n) = \left\{
\begin{array}{l l}
  1 & \quad \text{if} \quad \vert d_{1,2}(m,n)\vert \leq \tau(m,n) \\
     & \quad  (\text{or} \  0 < d_{1,2}(m,n) \leq \tau(m,n) \ \text{in the directed case}),\\
  0 & \quad \text{else}.\\
\end{array} \right.
\label{eq_sij}
\end{equation}
Now we can define the {\it event synchronization} $E$ between the two event series as the sum of $S(m,n)$
\begin{equation}
E = \sum_{m,n}S(m,n).
\label{eq_qij}
\end{equation}
This measure has the advantage that it can quantify interrelations between event-like time series and
that it allows for a flexible (dynamical) delay between the events. This is particularly different from 
the standard approach, where a considered lag (e.g., for cross-correlation) is constant and fixed at
each time point.

Applying the event synchronization approach, Eq.~(\ref{eq_qij}), for comparing spatially embedded
time series $x_i$ and $x_j$ at locations (nodes) $i$ and $j$ a network can be reconstructed in the
same way as in Eq.~(\ref{eq_thresholdC}). For selecting the threshold $T$ several approaches are possible.
One possibility is based on a significance test, where block bootstrapping can provide an empirical test 
distribution of the values of $\mathbf{E}$ and a preselected confidence level (e.g., 2\% or 5\%) provides the 
threshold $T$\cite{boers2014}. This 
procedure ensures that the network links represent only the strongest interrelations between the 
nodes.

Event synchronization based complex networks have been successfully used to investigate spatio-temporal
patterns during the Indian Summer monsoon \cite{malik2010,malik2012,stolbova2014}
or to study the origin and propagation of extreme rainfall in South America \cite{boers2013,boers2014}.
Although there is an obvious dominance of climate applications, this approach is also promising
for other fields, like plasma, turbulence, cardiological, or brain research.

\section{Developing a prediction scheme for extreme events}

\begin{figure}[htbp]
\begin{center}
\includegraphics[width=.7\columnwidth]{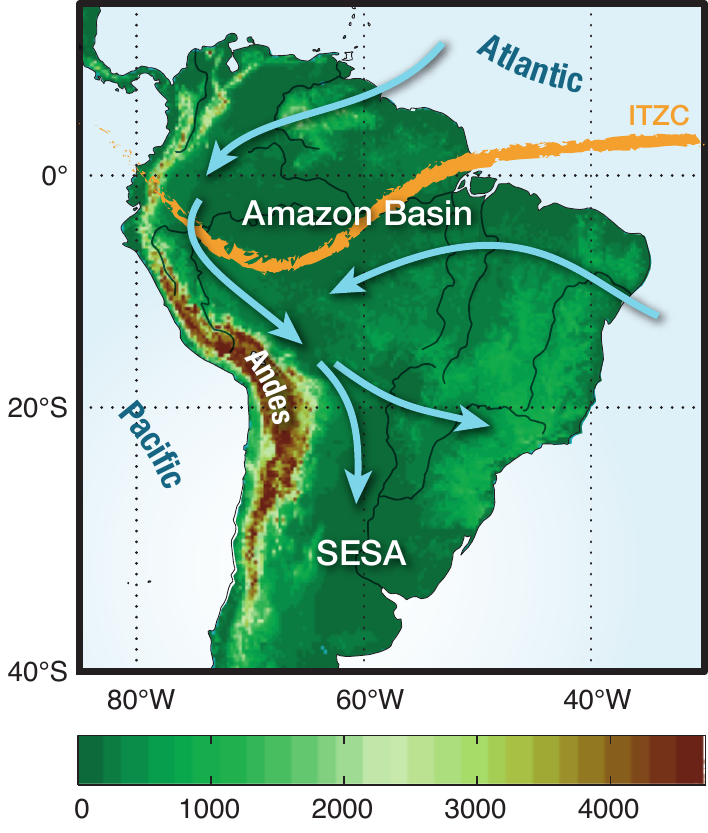}
\caption{Key features of the South American monsoon system. The blue arrows indicate
major moisture transport pathways.}
\label{climate_setting}
\end{center}
\end{figure}

To illustrate the potential of the complex network approach in the context of extreme climate events, we apply the
approach on South American extreme rainfall data and use the network topology for developing a prediction
scheme for extreme rainfall \cite{boers2014}. During the Australian summer season (December, January,
February), the differential heating between land and ocean amplifies the trade winds that enhances 
transport of moisture from the tropical Atlantic into the tropical Amazonian Basin and, thus, causes extended rainfall\cite{zhou1998}.
Due to evapotranspiration, wind, and the Andean orographic barrier, this water is first further transported westwards
and later, along the Andean mountain ridge, southwards towards the subtropics (Fig.~\ref{climate_setting}). Here, a frontal system converging
from the South and related with Rossby waves is responsible whether the moisture transport moves further 
eastward into the southeastern (SE) Brazil or towards the SE South America (SESA, central Argentinian plains)\cite{marengo2004}.
This variability of the exit moisture regions is also called South American rainfall dipole \cite{nogues1997}.

The spatio-temporal rainfall data used in our example is collected from the satellite-based Tropical Rainfall Measurement Mission 
(TRMM 3B42 V7, \cite{huffman2007}) with $3$-hourly temporal and $0.25^{\circ}\times0.25^{\circ}$ 
spatial resolution. In the following we will restrict the analysis to the Australian summer season 
and consider extreme rainfall events that are defined locally
by rainfall exceeding the $99$th percentile.

\begin{figure}[htbp]
\begin{center}
\includegraphics[width=.7\columnwidth]{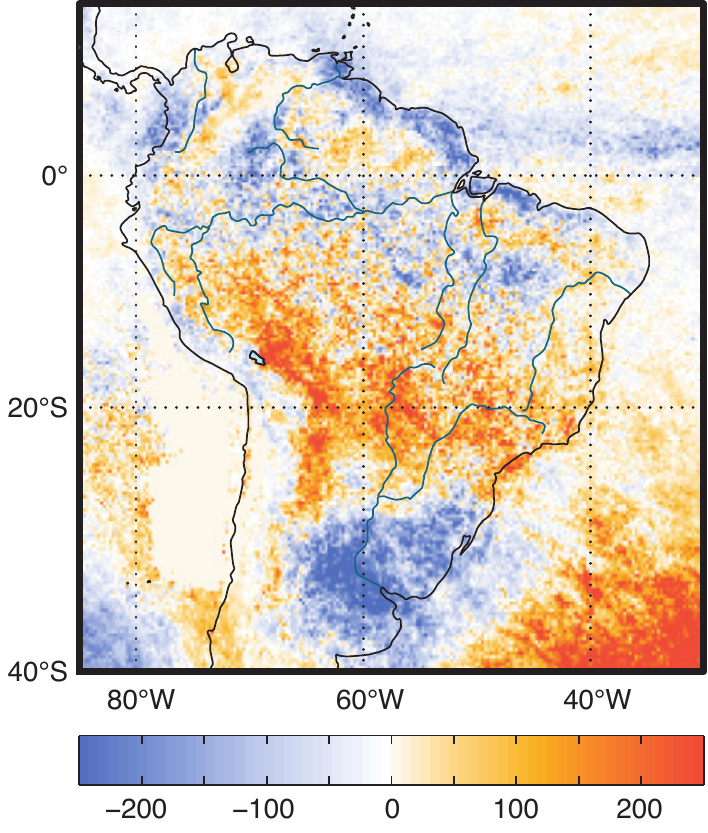}
\caption{Network divergence $\Delta\mathcal S$ of extreme rainfall network during Austral winter season.
Negative values indicate source and positive values sink regions of extreme rainfall.}
\label{SA_div}
\end{center}
\end{figure}

Using directed event synchronization, Eqs.~(\ref{eq_sij}--\ref{eq_qij}), of the rainfall extremes we 
reconstruct weighted and directed networks and calculate the in-strength $\mathcal S_{i}^{\text{in}}$
and out-strength $\mathcal S_{i}^{\text{out}}$, Eq.~(\ref{eq_in_outdegree}). Now we define the
{\it network divergence} as the difference between in- and out-strength:
\begin{equation}\label{eq_netwdiv}
\Delta\mathcal S_{i} = \mathcal S_{i}^{\text{in}} -  \mathcal S_{i}^{\text{out}}.
\end{equation}
Negative values of $\Delta\mathcal S$ indicate the source regions of extreme events whereas positive
values indicate sinks. Surprisingly, we find negative $\Delta\mathcal S$ values within the SESA region
(Fig.~\ref{SA_div}).
This means that this region is a source region of extreme rainfall although it is one of the
exit regions of the low-level moisture flow from the Amazon region. Now it would be interesting
to see to which other places the extreme rainfall from the SESA region will propagate.
For this purpose we consider the in-strength of all nodes conditioned by the source region SESA
and call it impact $\mathcal I_i(R)$ of region $R$ on node $i$:
\begin{equation}
\mathcal I_{i}(R) := \frac{1}{\vert R \vert}\sum_{j\in R}A_{ij}.
\end{equation}
$\vert R \vert$ is the number of nodes within the region of interest $R$ (here SESA). 
Casually speaking, $\mathcal I_i(R)$ measures the amount of extremes at site $i$ that have their origin in
region $R$. 

\begin{figure}[htbp]
\begin{center}
\includegraphics[width=.7\columnwidth]{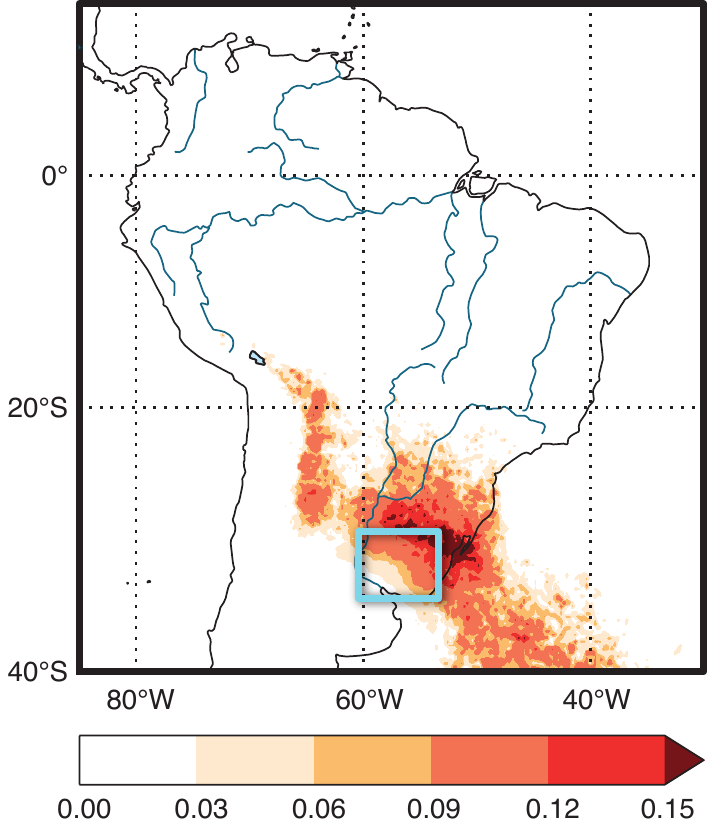}
\caption{Impact $\mathcal I_i(R)$ of a region (here SESA, marked by the box) in form of
contributing propagated extreme rainfall.}
\label{SA_impact}
\end{center}
\end{figure}

For the SESA region we find high values of $\mathcal I_i(\text{SESA})$ not only in the direct
vicinity of SESA but also at the eastern slopes of the Central Andes (Fig.~\ref{SA_impact}).
This result suggests that extreme rainfall at the Central (in particular Bolivian) Andes will precede after
rainfall events in the SESA region. 

The mechanism
behind this is an interplay between the orographic barrier, frontal systems approaching
from the South, and the southward moisture flow from the Amazon basin resulting
in the establishment of a wind channel attracting warm and moist air from the western Amazon region
into the SESA region\cite{boers2014}. Here it collides with the cold air of the frontal system from the South and
produces extended rainfall. This rainfall propagates together with the northern migration of
the frontal system and is bounded in the West by the Andean orography.

This fact can be used for defining a simple but very efficient prediction scheme as explained in detail by Boers 
et al.\cite{boers2014}
The precondition is a low-pressure anomaly in the SESA (geopotential height
anomaly $< -10$\,m). As long as this condition is fulfilled, during two days
extreme rainfall will appear at the eastern slopes of the Central Andes (in the 
range along the band that is marked by high values of $\mathcal I_i(\text{SESA}))$.
This rule allows positive prediction rates of 60\% and during El Ni\~no conditions
even of 90\%.

\section{Conclusion and Outlook}
In this paper we have presented an overview of a complex network based method for the analysis of
continuous dynamical systems. 
These methods are mainly basing on two concepts: (i) recurrence networks and (ii) event synchronization. 
The first one combines recurrence, a basic principle in dynamical systems, with complex networks. 
This way a rich variety of complex networks characteristics become available for time series analysis. 
Its potential have been demonstrated for the identification of sudden transitions from even short time series. 
The second one identifies events occurring almost synchronized in time in different spatial regions and uses 
then complex networks for the study of especially spreading and interactions of extreme events. 
We have uncovered with this technique a mechanism for the formation of extreme floods in the Andes which 
has led to a very efficient framework for predicting such extreme events. 
Additionally, this methodology can be used as a new tool for a critical comparison of different models of in particular natural systems.
Such complex network approaches have a strong potential for various fields, ranging from turbulence, via 
neuroscience and medicine to socio-economy.

However, there are several open problems to study in future. 
One direction is to extend these concepts to multivariate (spatio-temporal) data, e.g., different climatological or 
physiological parameters. 
A further challenge is the study of interacting systems of possible very different nature, e.g., climate and 
renewable energy generation or climate and health, from the network perspective. 
Another problem is a comprehensive mathematical foundation of these techniques including an appropriate test statistics.
Thus, we expect a pursuing and lively development and an increasing number of applications of these rather new concepts in the next future.

\section{Acknowledgement}
We would like to acknowledge support from the IRTG 1740/TRP 2011/50151-0, funded by the DFG/FAPESP, 
the DFG project ``Investigation of past and present climate dynamics and its stability by means of a 
spatio-temporal analysis of climate data using complex networks'' (MA 4759/4-1), and
project ``Gradual environmental change versus single catastrophe -- Identifying drivers of mammalian evolution''
(SAW-2013-IZW-2), funded by the Leibniz Association (WGL).
Moreover, we thank Niklas Boers for helpful comments.

\bibliographystyle{apsrev4-1}

\bibliography{rp,others,mybibs,geo}

\end{document}